\documentclass[usenatbib]{mn2e}
\usepackage{graphicx}
\usepackage{color}

\def\feii{Fe\,{\sevensize II}}
\def\feiiuv{Fe\,{\sevensize II}(UV)}
\def\feiiopt{Fe\,{\sevensize II}($\lambda$4570)}
\def\mgii{Mg\,{\sevensize II}}
\def\hbeta{H$\beta$}
\def\oi{O\,{\sevensize I}}
\def\oiii{[O\,{\sevensize III}]}
\def\caii{Ca\,{\sevensize II}}

\title[Implications from the FeII(opt)/FeII(UV) in quasars]
{Implications from the optical to UV flux ratio of FeII emission in
  quasars}
\author[H. Sameshima et al.]{H. Sameshima$^1$\thanks{E-mail:
    hsameshima@ioa.s.u-tokyo.ac.jp}, K. Kawara$^1$, Y. Matsuoka$^2$,
  S. Oyabu$^2$, N. Asami$^1$ and N. Ienaka$^1$ 
\\
$^1$Institute of Astronomy, University of Tokyo, 2-21-1, Osawa,
Mitaka, Tokyo 181-0015, Japan \\
$^2$Graduate School of Science, Nagoya University, Furo-cho,
Chikusa-ku, Nagoya 464-8602, Japan}

\begin{document}

\maketitle

\begin{abstract}
We investigate \feii\ emission in Broad Line Region (BLR) of AGNs by
analyzing the \feiiuv, \feiiopt\ and \mgii\ emission lines in 884
quasars in the Sloan Digital Sky Survey (SDSS) Quasar catalog in a
redshift range of $0.727 < z < 0.804$. \feiiopt/\feiiuv\ is used to
infer the column density of \feii-emitting clouds and explore the
excitation mechanism of \feii\ emission lines. As suggested before in
various works, the classical photoionization models fail to account
for \feiiopt/\feiiuv\ by a factor of 10, which may suggest anisotropy
of UV \feii\ emission; otherwise, an alternative heating mechanism
like shock is working. The column density distribution derived from
\feiiopt/\feiiuv\ indicates that radiation pressure plays an important
role in BLR gas dynamics. We find a positive correlation between
\feiiopt/\feiiuv\ and the Eddington ratio. We also find that almost
all \feii-emitting clouds are to be under super-Eddington conditions
unless ionizing photon fraction is much smaller than that previously
suggested. Finally we propose a physical interpretation of a striking
set of correlations between various emission-line properties, known as
``Eigenvector 1''.

\end{abstract}


\section{Introduction}

According to the models of nucleosynthesis, much of Fe comes from Type
Ia supernovae (SNe Ia), while $\alpha$ elements such as O and Mg come
from Type II supernovae (SNe II). Because of the difference in
lifetime of the progenitors, Fe-enrichment delays relative to $\alpha$
elements. Hence the abundance ratio of Fe to $\alpha$ elements
[Fe/$\alpha$] should have a sudden break at 1--2 Gyr after the initial
burst of star formation (\citealt{hf}; \citealt*{ysi96};
\citealt*{ysi98}, but see also recent studies, such as \citealt{matt}
and \citealt{tot}, indicating a significant number of SNe Ia on
relatively short timescales). Under the assumption that the
\feii/\mgii\ flux ratio reflects [Fe/Mg], various groups have measured
\feii/\mgii\ in high-redshift quasars hoping to discover such a break
(e.g., \citealt*{eth}; \citealt{kwr}; \citealt{die02},
\citeyear{die03}; \citealt{iwa02}, \citeyear{iwa04}; \citealt*{fck};
\citealt{mai}; \citealt{kurk}; \citealt{sam}). However, these efforts
have ended up with finding a large scatter of \feii/\mgii\ showing
little evolution. 

A doubt is, thus, cast on the assumption that \feii/\mgii\  reflects
[Fe/Mg]. For examples, \citet{ver03} suggested that \feii/\mgii\
depends not only on the abundance but also on the microturbulence of
\feii-emitting clouds. \citet{tzk} showed that \feii/\mgii\ correlates
with the X-ray photon index, the full width at half maximum (FWHM) of
\mgii, the black hole mass, etc. For these reasons, prior to deriving
the abundance from \feii/\mgii, we must first clarify these
non-abundance effects on the \feii\ emission as well as the source of
the \feii\ excitation.

Column densities of clouds, which are considered to be one of the
non-abundance factors largely affecting the \feii\ emission, have
lately attracted attention for their significances in determining
whether or not the radiation pressure plays an important role in BLR
gas dynamics. \citet{mar08} considered the effect of radiation
pressure from ionizing photons on estimating of the black hole mass,
which is based on the application of the virial theorem to broad
emission lines in AGN spectra, and suggested that the black hole mass
can be severely underestimated if the effect of radiation pressure is
ignored. \citet{net} then used the $M_{BH}-\sigma_*$ relation for a
test of this suggestion, where $M_{BH}$ is the black hole mass and
$\sigma_*$ is the velocity dispersion of host galaxies, concluding
radiation pressure effect is unimportant, while \citet{mar09} found
the importance of radiation pressure by taking into account the
intrinsic dispersion associated with the related parameters, in
particular, column densities of BLR clouds. However, there are no
reliable column density estimates from observations up to date.

In this paper, we will estimate column densities of quasars selected
from the SDSS using \feii, for investigating the excitation mechanism
of \feii\ emission and the BLR gas dynamics. In section 2, we perform
numerical calculations of \feii\ emission lines to establish a method
for estimating column densities. In section 3, \feii\ emission lines
in the UV and optical as well as \mgii\ emission lines are measured in
the SDSS quasars. The results and discussion about \feii\ emission
mechanism, radiation pressure, and the variety of quasar spectra
called as ``Eigenvector 1'' are given in section 4. Throughout this
paper, we assume a cosmology with $\Omega_m=0.3$,
$\Omega_{\Lambda}=0.7$ and $H_0=70\ \mathrm{km\ s^{-1}\ Mpc^{-1}}$.

\section{Methodology}

In the following, we use \feiiuv\ to denote the UV \feii\ emission
lines in $2000 < \lambda < 3000$ \AA, \feiiopt\ to the optical \feii\
emission lines in $4435 < \lambda < 4685$ \AA, and \mgii\ to the
\mgii\ $\lambda 2798$ emission line.

\subsection{\feiiopt/\feiiuv\ to measure column densities}

\begin{figure}
  \includegraphics[width=84mm]{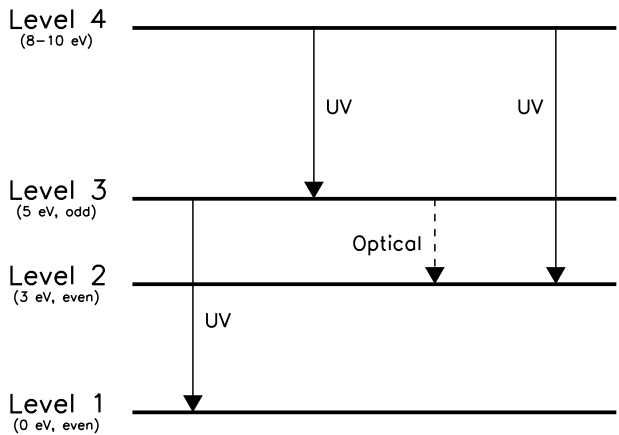}
  \caption{
    Simplified \feii\ Grotrian diagram. Note that each Level
    represents a large number of levels that have nearly the same
    excitation energies. {\it Solid arrow} indicates UV \feii\
    emission lines, while {\it dashed arrow} indicates optical \feii\
    emission lines.
  }
\label{fig:grot_diag}
\end{figure}

Figure \ref{fig:grot_diag} shows a simplified \feii\ Grotrian
diagram. As can be seen, Level 3$-$Level 2 transitions give rise to
optical \feii\ emission lines such as the \feiiopt\ bump. Branching
ratio of Level 3$-$Level 1 UV resonance transitions is significantly
higher than that of Level 3$-$Level 2 optical transitions. Hence
strong optical \feii\ emission requires a large optical depth between
Level 1 and Level 3 such as $\tau_{13} > 10^3$ in order to transform
UV \feii\ lines to optical \feii\ lines through a large number of
scatterings (cf. \citealt{cj}). Thus \feiiopt/\feiiuv\ flux ratio must
strongly depend on $\tau_{13}$ and $\tau_{23}$, which are the optical
depths for photons emitted through Level 3$-$Level 1 and Level
3$-$Level 2 transitions, respectively. If so, the \feiiopt/\feiiuv\
can be an indicator of the column density.

Here we use a quite simple model to indicate the dependence of
\feiiopt/\feiiuv\ on the column density. First, we ignore the Level 4
in Figure \ref{fig:grot_diag} and consider the \feii\ as three-level
system. Second, we assume thermal equilibrium population between Level
1 and Level 2. Third, we assume an expression of the local escape
probability given by \citet{nw} as
$\epsilon_{ij}=(1-\tau_{ij})/\tau_{ij}$. Although the model adopting
these assumptions is obviously too oversimplified, it is useful to
qualitatively understand how the line ratio depends on the physical
parameters. The flux ratio is then written as 
\begin{eqnarray}
  \frac{\mbox{\feiiopt}}{\mbox{\feiiuv}} &\approx& \frac{n_3 A_{32}
    \epsilon_{32} h\nu_{23}}{n_3 A_{31} \epsilon_{31} h\nu_{13}} \\
  &\propto& \exp \left( \frac{2.9\ \mathrm{eV}}{kT} \right)
  \frac{1-e^{-\tau_{23}}}{1-e^{-\tau_{13}}} \label{eq:fefe_rough}
\end{eqnarray}
where $n_3$ is the population of level 3, and $A_{ij}$ is the
spontaneous emission rate from the level $i$ to the level
$j$. Following the fact that $\tau_{13} \gg 1$, the ratio can be
roughly reduced to \feiiopt/\feiiuv\ $\propto 1-e^{-\tau_{23}}$. Since
$\tau_{23}$ is proportional to the column density, the ratio is an
increasing function of the column density. We will discuss further on
this matter in the following.

\subsection{Photoionization models in the LOC scenario}
Here we will show more sophisticated model calculations of
\feiiopt/\feiiuv. We performed photoionization model calculations with
version C06.02 of the spectral simulation code Cloudy, last described
by \citet{fer98}, combined with a 371 level Fe$^+$ model (up to
$\sim$11.6 eV, \citealt{ver99}). The incident continuum is defined as
\begin{equation}
  f_\nu \propto
  \nu^{\alpha_{UV}}\exp(-h\nu/kT_{BB})\exp(-kT_{IR}/h\nu) +
  a\nu^{\alpha_{X}} \label{eq:sed}
\end{equation}
The first term in the right-hand side of the equation (\ref{eq:sed})
expresses an accretion disk component, which is usually called {\it
  Big Bump}. The $kT_{BB}$ and the $kT_{IR}$ indicate the higher and
the lower cut off energies, respectively. The second term expresses a
power-law X-ray component, which is set to zero below 1.36 eV, while
to fall off as $\nu^{-3}$ above 100 keV. The coefficient $a$ is set to
produce the optical to X-ray spectral index $\alpha_{ox}$
\footnote{The optical to X-ray spectral index $\alpha_{ox}$ is defined
  as $f_\nu(2$ keV$)/f_\nu(2500$\AA$)=403.3^{\alpha_{ox}}$}. We set
these parameters to $(\alpha_{UV},\ \alpha_{X},\ \alpha_{ox},\
T_{BB},\ kT_{IR})=(-0.2,\ -1.8,\ -1.4,\ 1.5\times 10^5\ \mathrm{K},\
0.136\ \mathrm{eV})$, which are adopted in \citet{tzk}. This incident
continuum illuminates a single cloud with hydrogen density $n_H$,
ionization parameter $U$($\equiv \Phi/n_H c$, where $\Phi$ is the
ionizing photon flux, $c$ is the velocity of light), column density
$N_H$ and solar abundance. We calculated the models in a range of
$N_H=10^{21}-10^{25}\ \mathrm{cm}^{-2},\ n_H=10^7-10^{14}\
\mathrm{cm}^3$ and $U=10^{-5}-10^{0}$.

In locally optimally emitting clouds (LOCs; \citealt{bal95}), each
line is emitted from clouds with a wide range of gas hydrogen density
and distance from the central continuum source, and the observed
spectra is reproduced by integrating these clouds with an appropriate
covering fraction distribution. The observed emission line flux is,
then, expressed as: 
\begin{equation}
  L_{line} \propto \int\!\!\!\int r^2 F(r,n_H)f(r)g(n_H)dn_H
  dr \label{eq:loc}
\end{equation}
where $F(r,n_H)$ is the emission line flux of a single cloud at a
distance $r$ from the central continuum source and with $n_H$, $f(r)$
is a cloud covering fraction with distance $r$, and $g(n_H)$ is a
fraction of clouds with $n_H$. \citet{mat07} showed that \oi\ and
\caii\ emission lines in quasars, which are likely to emerge from the
same gas as the \feii\ emission lines, are well reproduced by a LOC
model with $f(r) \propto r^{-1}$ and $g(n_H) \propto n_H^{-1}$, hence
we have adopted these covering distributions. \citet{bal95} suggested
that since clouds at large distances from the continuum source will
form graphite grains which heavily suppress the line emissivity,
clouds with $\Phi < 10^{18}\ \mathrm{s^{-1}cm^{-2}}$ must be excluded
from integration. Thus we have calculated equation (\ref{eq:loc}) for
clouds which correspond $\Phi \ge 10^{18}\ \mathrm{s^{-1}cm^{-2}}$.

\begin{figure}
  \includegraphics[width=84mm]{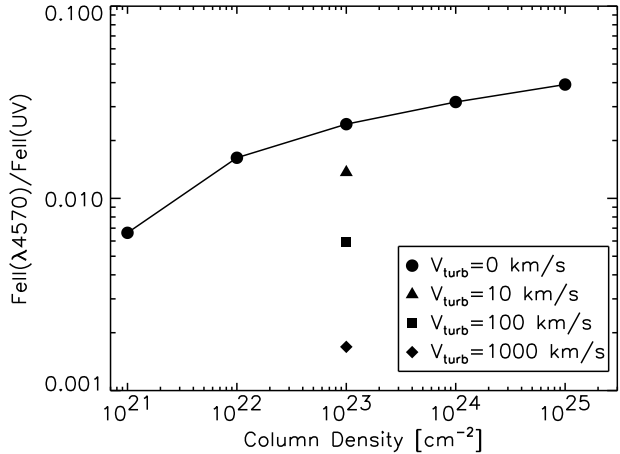}
  \caption{
    \feiiopt/\feiiuv\ vs. column densities based on the
    photoionization models. $V_{turb}$ is the microturbulence velocity
    of clouds. Note that, in the LOC, the calculation results are not
    dependent on $n_H$ and $U$ since clouds with different hydrogen
    density and distance  from central continuum source are all
    integrated.
  }
\label{fig:fefe_phot}
\end{figure}

Figure \ref{fig:fefe_phot} illustrates \feiiopt/\feiiuv\ as a function
of the column density for photoionized clouds. \feiiopt/\feiiuv\
increases as the column density increases. It also shows that
\feiiopt/\feiiuv\ decreases as the microturbulent velocity increases,
consistent with \citet{ver03}, and that \feiiopt/\feiiuv\ is maximum
when no microturbulence is assumed to exist. This is explained as
follows. Increasing microturbulent velocities broadens the line
absorption profile, resulting in enhancement of the continuum
photoexcitation. This effect is relatively large for high energy
levels where the collisional excitation is inefficient for their high
excitation potential. Thus large microturbulent velocities relatively
enhances the continuum photoexcitation to the Level 4 in Figure
\ref{fig:grot_diag}, leading to emit more fluxes in the UV \feii\
emission lines, thus decreasing \feiiopt/\feiiuv. For photoionized
clouds, \feiiopt/\feiiuv\ can be used as a column density indicator
unless microturbulent velocities vary much from cloud to cloud.

It is noted that calculations adopting the other shapes of the
incident continuum\footnote{Two types of SED given by \citet*{nmt} are
  adopted. One is set to reproduce the ordinary SED of broad-line
  Seyfert 1 galaxies, and the other is set to reproduce that of
  narrow-line Seyfert 1 galaxies.} show little changes for
\feiiopt/\feiiuv, indicating little dependence on the spectral energy
distribution (SED) of ionizing continuum.

\subsection{Collisionally ionized models}
As an alternative to photoionization models, we consider models in
which a cloud is assumed to be in collisional equilibrium at a given
electron temperature $T_e$, and call them {\it collisionally ionized
  models}. In these models, the specific heating mechanism is not
accounted and arbitrary electron temperatures are given. It is noted
that \feii\ is collisionally excited in either mechanically heated
clouds such as through shocks or photoionized clouds. In the latter
case, the heating mechanism is specified to be the eating of the
incident UV and X-ray photons.

\citet{joly} offers collisionally ionized model calculations. In her
models, \feii\ is approximated by a 14-level (up to $\sim$5.7 eV), and
the emission region is assumed to be a homogeneous slab with constant
hydrogen density $n_H$, column density $N_H$, electron temperature
$T_e$, and not to receive any external radiation. The electron
temperature ranges from 6,000 K to 15,000 K, the density from
$10^{10}$ to $10^{12}\ \mathrm{cm}^{-3}$ and the column density from
$10^{21}$ to $10^{25}\ \mathrm{cm}^{-2}$.

\begin{figure}
  \includegraphics[width=84mm]{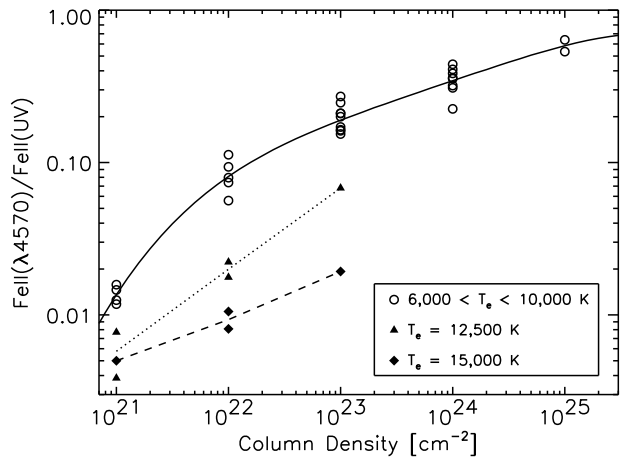}
  \caption{
    \feiiopt/\feiiuv\ vs. column densities based on the collisionally
    ionized models taken from \citet{joly}. The density ranges from
    $10^{10}$ to $10^{12}\ \mathrm{cm^{-3}}$ and the column density
    from $10^{21}$ to $10^{25}\ \mathrm{cm^{-2}}$. {\it Circles}
    indicate the models with $6,000 < T_e < 10,000$ K, {\it triangles}
    with $T_e = 12,500$ K and {\it diamonds} with $T_e = 15,000$
    K. The solid line indicates the fitted 4th-order polynomial for
    samples with $6,000 < T_e < 10,000$ K. The dotted and the dashed
    lines connect the median values of \feiiopt/\feiiuv\ at each
    column density for samples with $T_e = 12,500$ K and $T_e =
    15,000$ K, respectively.
  }
\label{fig:fefe_col}
\end{figure}

Figure \ref{fig:fefe_col} shows \feiiopt/\feiiuv\ as a function of the
column density, taken from \citet{joly}. As expected from equation
(\ref{eq:fefe_rough}), it can be seen that \feiiopt/\feiiuv\ increases
with the column density while decreases with temperature. However,
Figure \ref{fig:fefe_col} shows that \feiiopt/\feiiuv\ does not so
much depend on the temperature in $6,000 < T_e < 10,000$ K. It is
noted that \citet{col} and \citet{joly} showed emission line ratios
including \feii\ in quasars are well accounted by cold clouds with
$6,000 < T_e < 10,000$ K. Thus it is reasonable to assume that
\feii-emitting clouds have the temperatures in a range of $6,000 < T_e
< 10,000$ K, and the \feiiopt/\feiiuv\ depends little on the
temperature while strongly varies with the column density. We fit the
data of $6,000 < T_e < 10,000$ K models by a 4th-order polynomial, and
find 
\begin{equation}
  y = -0.72 + 0.28x - 0.046x^2 + 0.031x^3 - 0.0089x^4 \label{eq:polyfit}
\end{equation}
where $y=\log$ \feiiopt/\feiiuv\ and $x=\log N_{H} [\mathrm{cm^{-2}}]
- 23$. We will use this relation to estimate the column density of
quasars.

\section{Analysis of quasar spectra}
\subsection{Sample selection} \label{sec:sample}
We have analyzed quasar spectra selected from the fourth edition of
the SDSS Quasar catalog (\citealt{snd}). SDSS uses a dedicated 2.5 m
telescope at the Apache Point Observatory equipped with a CCD camera
to image the sky in five optical bands, and two digital spectrographs,
one covering a wavelength range of 3800\AA\ to 6150\AA\ and the other
from 5800\AA\ to 9200\AA. The spectral resolution ranges from 1850 to
2200. The fourth edition of the SDSS Quasar catalog consists of the
objects in the Fifth Data Release, and contains 77,429 quasars.

In order to measure the \feiiuv\ and \feiiopt\ emission lines
simultaneously, we have selected spectra which cover the wavelength
range of 2200 to 5100 \AA\ in the rest frame, corresponding to the
redshift range from 0.727 to 0.804. 2,189 objects meet this
requirement. Then we have checked the signal to noise ratio (S/N) per
pixel for each spectrum which fulfills median S/N $>$ 10 per pixel at
their continuum levels for accurate flux measurements. 946 objects meet
this requirement. All the spectra were inspected by eyes and 62
spectra were rejected because of wavelength discontinuity, terrible
contamination by host galaxy star light, etc. Our final sample thus
consists of 884 spectra.

Prior to the measurements, the quasar spectra were dereddened for the
Galactic extinction according to the dust map by \citet*{slg} using
the Milky Way extinction curve by \citet{pei}.

\begin{figure}
  \includegraphics[width=84mm]{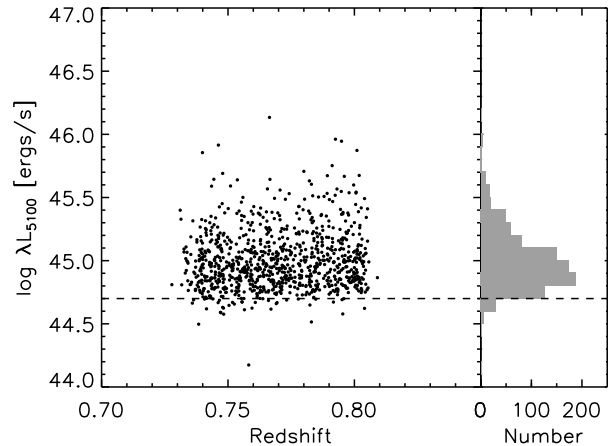}
  \caption{
    Luminosity $\lambda L_{5100}$ vs. redshift for our
    sample. Luminosity distribution is also displayed in the right
    panel. The dashed line shows a rough estimate of minimum quasar
    luminosity for our sample.
  }
\label{fig:lumi_limit}
\end{figure}

Since SDSS quasars are flux-limited in the survey, low luminosity
quasars are lost in the SDSS sample. Figure \ref{fig:lumi_limit} shows
luminosity versus redshift for our sample. Because of the narrow
redshift coverage for our sample, the minimum luminosity is almost
constant and roughly estimated to be $\lambda L_{5100}=10^{44.7}$
ergs/s.

\subsection{Continuum and line fitting} \label{sec:fit}
In the UV to optical, the quasar continuum is composed of (i) the
power-law continuum $F_\lambda^{PL}$, (ii) the Balmer continuum
$F_\lambda^{BaC}$ and (iii) the \feii\ pseudo-continuum
$F_\lambda^{FeII}$. Thus we assumed a following formula as a model
continuum $F_\lambda^{cont}$: 
\begin{equation}
  F_\lambda^{cont} = F_\lambda^{PL} + F_{\lambda}^{BaC} + F_\lambda^{FeII}
\end{equation}

\subsubsection{Power-Law continuum}
The power-law continuum is simply written as follows:
\begin{equation}
  F_\lambda^{PL} = F_{5100} \left( \frac{\lambda}{5100} \right)^\alpha
\end{equation}
The free parameters of this model are a scaling factor $F_{5100}$ and
a power-law index $\alpha$. We chose three fitting ranges, 2200-2230
\AA, 4180-4220 \AA\ and 5050-5100 \AA, as continuum windows, since
these area have little emission lines (see Figure
\ref{fig:cont_fit}). There are, however, the Balmer continuum and the
\feii\ pseudo-continuum underneath these regions, requiring some
corrections.

\citet{tzk} gives 14 quasar spectra covering a wide wavelength range
and measured accurately their continuum levels. We fitted power-law
continuum models to their spectra in the continuum windows, and
compared the continuum levels with those given by \citet{tzk}. We
found that our method systematically overestimates the continuum
levels, 10.1\% at 2200-2230 \AA, 5.7\% at 4180-4220 \AA\ and 3.4\% at
5050-5100 \AA. According to these results, we first reduced the flux
densities of the object by these amount at each continuum window, then
fitted the power-law continuum. An example of the fitted power-law
continuum is indicated as the dashed line in Figure
\ref{fig:cont_fit}. The measurement error of the continuum levels is
estimated to be less than 10\%.

\subsubsection{Balmer continuum}
\citet{gra} gives a formula describing the Balmer continuum produced
by a uniform temperature, partially optically thick cloud: 
\begin{equation}
  F_\lambda^{BaC} = F_{BaC} B_\lambda(T_e) \left[ 1-\exp \left\{
      -\tau_{BE} \left( \frac{\lambda}{\lambda_{BE}} \right)^3
    \right\} \right] \label{eq:bac}
\end{equation}
where $B_\lambda(T_e)$ is the Planck function at the electron
temperature $T_e$, and $\tau_{BE}$ is the optical depth at the Balmer
edge at $\lambda = 3646$ \AA. \citet{kurk} assumed gas clouds of
uniform temperature ($T_e=15,000$ K) and the optical depth fixed to
$\tau_{BE}=1$, and fit equation (\ref{eq:bac}) to their sample quasar
spectra to estimate the strength of the Balmer continuum (see also
\citealt{die03}). We followed their method and assumed $T_e=15,000$ K,
$\tau_{BE}=1$. The only one parameter, namely the scale factor
$F_{BaC}$, is set free and is decided by fitting equation
(\ref{eq:bac}) to the power-law subtracted spectrum at 3600$-$3645
\AA. An example of the fitted Balmer continuum is indicated as the
dotted line in Figure \ref{fig:cont_fit}.

\subsubsection{\feii\ pseudo-continuum}
Since \feii\ has enormous energy levels, neighboring emission lines
contaminate heavily with each other, which makes it difficult to
measure the \feii\ emission lines. One approach to measure the \feii\
emission lines is to use \feii\ templates. So far, several \feii\
templates are derived from the narrow-line Seyfert 1 galaxy, I Zw 1.

In the UV, \citet{vw} and \citet{tzk} give their \feii\ templates. The
template given by \citet{vw} do not cover around \mgii\
line. \citet{tzk} used a synthetic spectrum calculated with the Cloudy
photoionization code in order to separate the \feii\ emission from the
\mgii\ line, and derived semiempirically the \feii\ template which
covers around the \mgii\ line. Since we want to measure the \mgii\
emission line, we decided to use the UV \feii\ template given by
\citet{tzk}.

In the optical, \citet*{vjv} and \citet{tzk} open their \feii\
templates to the public. \citet{vjv} carefully analyzed the \feii\
emission lines in I Zw 1, finding that the \feii\ lines are emitted
from both BLR and Narrow Line Region (NLR). They succeeded to separate
them and called the broad line system L1 and the narrow line system
N3, respectively. \citet{tzk} also analyzed the spectrum of I Zw 1 and
derived the optical \feii\ template, which was however not separated
into the BLR and the NLR components. We applied both the broad line
system L1 template given by \citet{vjv} and the optical \feii\
template given by \citet{tzk} to all of our samples, finding that the
latter has a slightly smaller average $\chi_\nu^2$ value (median
$\chi_\nu^2 \sim 1.48$ for \citealt{tzk}, while median $\chi_\nu^2
\sim 1.58$ for \citealt{vjv}). Here we adopt to use the optical \feii\
template given by \citet{tzk}.

\begin{figure}
  \includegraphics[width=84mm]{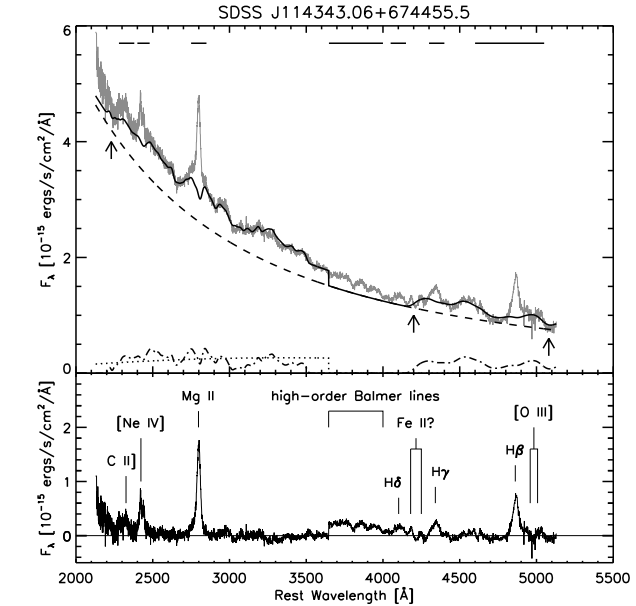}
  \caption{
    {\it Top:} Sample quasar spectrum ({\it gray line}) with the
    fitted power-law continuum $F_\lambda^{PL}$ ({\it dashed line}),
    the Balmer continuum $F_\lambda^{BaC}$ ({\it dotted line}), the
    \feii\ pseudo-continuum $F_\lambda^{FeII}$ ({\it dot-dashed line})
    and the sum of the three continua $F_\lambda^{cont}$  ({\it solid
      line}). Arrows indicate the continuum windows adopted in the
    power-low continuum fitting. Horizontal thick bars indicate the
    masked region adopted in the \feii\ pseudo-continuum fitting. {\it
      Bottom:} The continuum subtracted spectrum.
  }
\label{fig:cont_fit}
\end{figure}

Prior to applying the \feii\ template to each quasar, broadening of
the template spectrum is needed. Thus we modeled the \feii\ flux
density as follows:
\newpage
\begin{eqnarray}
  F_\lambda^{FeII}(x) &=& F_{FeII} \int_{-\infty}^{\infty}
  F_\lambda^{template}(x^{\prime}) \nonumber \\ 
  && \times \exp\left[ -\frac{4c^2\ln
      2(x-x^{\prime})^2}{{\mbox{FWHM}_{conv}}^2} \right]
  dx^{\prime} \label{eq:feiiconv}
\end{eqnarray}
where $x \equiv \ln \lambda$, $c$ is the velocity of light and
$\mbox{FWHM}_{conv}$ represents the FWHM of the convolved Gauss
function. We first calculated the equation (\ref{eq:feiiconv}) for
$\mbox{FWHM}_{conv}=0-5000$ km/s stepped by 100 km/s, thus prepared 51
\feii\ emission line models. Then we flux-scaled each model to fit the
continuum subtracted spectrum (i.e., the spectrum after subtracting
the power-law and the Balmer continuum), and adopted the model which
gives the smallest $\chi_\nu^2$ value. We decided mask regions as
follows; 2280-2380\AA\ for C\,{\sevensize II}] $\lambda$2326;
2400-2480\AA\ for [Ne\,{\sevensize IV}] $\lambda$2423 and
Fe\,{\sevensize III}; 2750-2850\AA\ for \mgii; 3647-4000\AA\ for
high-order Balmer lines; 4050-4150\AA\ for H$\delta$; 4300-4400\AA\
for H$\gamma$; 4600-5050\AA\ for He\,{\sevensize II} $\lambda$4686,
\hbeta\ and \oiii\ $\lambda\lambda$4959,5009. An example of the fitted
\feii\ pseudo-continuum is indicated as the dot-dashed line in Figure
\ref{fig:cont_fit}.

\subsubsection{\mgii\ lines}
\begin{figure}
  \includegraphics[width=84mm]{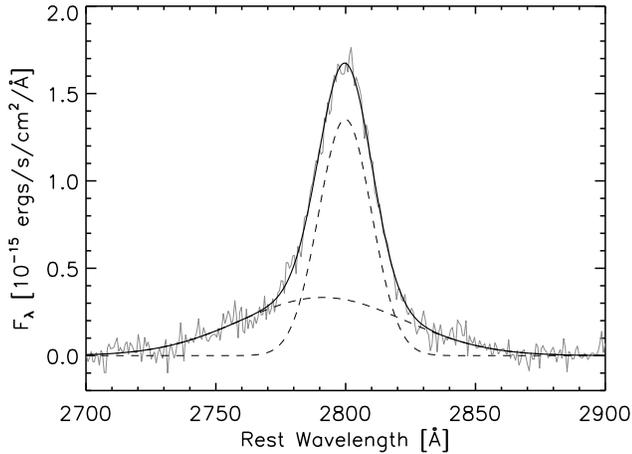}
  \caption{
    \mgii\ emission line ({\it gray line}) with fitted two gaussian
    components ({\it dashed lines}) and the sum of the two gaussians
    ({\it solid line}).
  }
\label{fig:mg_fit}
\end{figure}

After subtracting the continuum components, we have measured the
\mgii\ emission line for estimating the black hole mass from its
FWHM. We have fitted the \mgii\ emission line profile by two gaussian
components. Figure \ref{fig:mg_fit} shows an example of the \mgii\
emission line fitting.

\subsection{Black hole mass and Eddington luminosity}
For the classical black hole mass estimate in which the radiation
pressure effect is neglected, we use the following formula given by
\citet{mj}: 
\begin{equation}
  M_{BH,0} = 3.37 \left( \frac{\lambda L_{3000}}{10^{37}\ \mathrm{W}}
  \right)^{0.47} \left[ \frac{\mbox{FWHM(\mgii)}}{\mathrm{km\ s^{-1}}}
  \right]^{2} M_\odot \label{eq:bh_mj}
\end{equation}
The classical Eddington luminosity is given as follows:
\begin{equation}
  L_{Edd,0}=\frac{4\pi cGM_{BH,0}m_p}{\sigma_T} \sim 1.26 \times
  10^{38} \frac{M_{BH,0}}{M_{\odot}} \ [\mathrm{ergs/s}] 
\end{equation}
where $c$ is the velocity of light, $m_p$ is the proton mass, and
$\sigma_T$ is the Thomson cross-section.

On the other hand, \citet{mar08} suggested that the force of the
radiation pressure should be corrected to derive the black hole
mass. They give the radiation pressure corrected black hole mass
$M_{BH,rad}$ and Eddington luminosity $L_{Edd,rad}$ as follows:
\begin{equation}
M_{BH,rad} = M_{BH,0} + \frac{L_{bol}}{L_{Edd,0}} \left(
  1-a+\frac{a}{\sigma_T N_H} \right) M_{BH,0} \label{eq:bh_rad}
\end{equation}
\begin{equation}
L_{Edd,rad} = \frac{L_{Edd,0}}{1-a+a/(\sigma_T
  N_H)} \label{eq:edd_rad}
\end{equation}
\begin{equation}
a \equiv \frac{L_{ion}}{L_{bol}}
\end{equation}
where $L_{bol}$ is the bolometric luminosity, $L_{ion}$ is the total
luminosity of the ionizing continuum (i.e., $h\nu >$ 13.6 eV), and $a$
is the ionizing photon fraction. The second term in the right hand
side of equation (\ref{eq:bh_rad}) represents the correction term of
the radiation pressure. We here adopt a bolometric correction
$L_{bol}=9\lambda L_{5100}$ given by \citet{kas}.

\subsection{Error estimate}
We performed a Monte-Carlo simulation similar to those done in
\citet{hu} for estimating the measurement errors. The detail of the
procedure is as follows.

(i) {\it Generating a composite spectrum}.
Following \citet{van}, we generated a composite spectrum using all
of our samples. This composite spectrum represents a typical quasar
spectrum for our samples.
(ii) {\it Obtaining typical emission line profiles}.
We applied the measurement methods written in \S
\ref{sec:fit} to the composite spectrum, and
obtained typical emission line profiles for \feiiuv, \feiiopt\ and
\mgii.
(iii) {\it Making artificial spectra}.
We combined these line profiles with the power-low continuum and the
Balmer continuum. Thus the simulated spectrum is written as follows.
\begin{eqnarray}
  F_\lambda^{sim} &=& F_\lambda^{PL}(F_{5100},\alpha) +
  F_\lambda^{BaC}(F_{BaC}) \nonumber \\
  &+& F_\lambda^{FeII_{UV}}(EW_{FeII_{UV}}) \nonumber \\
  &+& F_\lambda^{FeII_{\lambda4570}}(EW_{FeII_{\lambda4570}}) \nonumber \\
  &+& F_\lambda^{MgII}(EW_{MgII},FWHM_{MgII}) \label{eq:simflux}
\end{eqnarray}
Note that, for simplicity, we ignored the broadening of the
pseudo-\feii\ continuum. Values of input parameters, which are given
in the parentheses in equation (\ref{eq:simflux}), are randomly
sampled from probability distributions that are made to reflect the
observations. Thus, we generated 1,000 simulated spectra.
(iv) {\it Generating a noise template}.
Using all the noise spectra produced by the SDSS pipeline for our
samples, we generated a composite spectrum following \citet{van} and
named it a noise template. This noise template is scaled so that the
resulting median S/N to be 10 per pixel at the continuum level for
each simulated spectrum, and is treated as its noise.

Now we have the 1,000 simulated spectra with their noise. The
measurement methods written in \S \ref{sec:fit} are applied to these
simulated spectra. We calculate the value $\delta_{sim}=(P_{out} -
P_{in})/P_{in}$ for each simulated spectrum where $P_{in}$ represents
the input parameters (i.e., the values given in the parentheses in
equation (\ref{eq:simflux})) and $P_{out}$ represents the
corresponding measured values for the simulated spectra. We regard
$\sigma_{sim}$, a standard deviation of $\delta_{sim}$, as $1\sigma$
error of the measurement. Thus we evaluate the measurement errors to
be 16.4\% for EW of \feiiuv, 22.9\% for EW of \feiiopt\ and 7.9\% for
FWHM(\mgii). The simulation implies the measurement error to be 2.9\%
for the luminosity, which is less than 10\% estimated in the power-law
fitting. Therefore we decided to evaluate the measurement error to be
10\% for $\lambda L_{3000}$ and $\lambda L_{5100}$.

\section{Results and Discussion}

\subsection{On the excitation mechanism of \feii\ emission}
\begin{figure}
  \includegraphics[width=84mm]{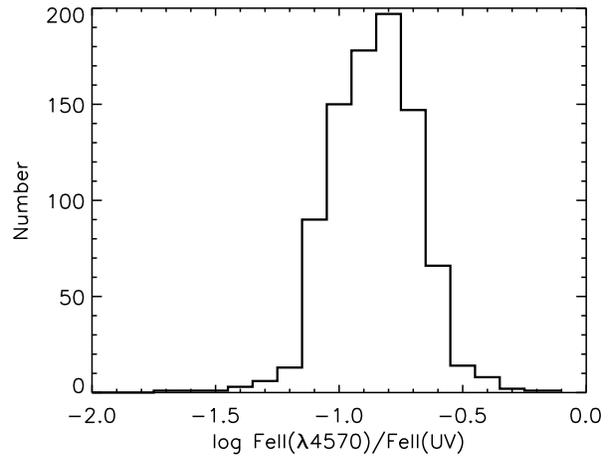}
  \caption{
    Observed \feiiopt/\feiiuv\ flux ratio distribution. The average of
    $\log$ \feiiopt/\feiiuv\ is $-0.8$ and the standard deviation is
    0.2 dex, in disagreement with the range that the photoionization
    models predict.
  }
\label{fig:fefe_dist}
\end{figure}

Figure \ref{fig:fefe_dist} shows the observed \feiiopt/\feiiuv\
distribution. As can be seen from the comparison between this and
Figure \ref{fig:fefe_phot}, our photoionization models underpredict
the \feiiopt/\feiiuv\ by a factor of 10, failing to account for the
observations. This result is consistent with the preceding study by
\citet{bal04}. Additional microturbulence to the photoionized clouds
gets the situation even worse. Thus, Figure \ref{fig:fefe_dist} seems
to challenge classical photoionized pictures of \feii-emitting
clouds. On the other hand, in the case of the collisionally ionized
models shown in Figure \ref{fig:fefe_col}, the observed
\feiiopt/\feiiuv\ flux ratios are well reproduced with $10^{22} < N_H
< 10^{24}\ \mathrm{cm^{-2}}$ and with $6,000 < T_e < 10,000$ K.

These results give two remarks: (1) the \feii-emitting clouds in
quasars are heated to $6,000 < T_e < 10,000$ K; (2) the UV and the
X-ray photons, which are the heating source in our photoionization
models, fail to heat the gas to such temperatures (probably heat the
gas too hot!). One possible interpretation is that the \feii-emitting
clouds are heated by an alternative mechanism such as through
shocks. Here we note that there is a reverberation mapping study
implying shock heating for \feii\ emission. \citet{kue} analyzed
optical \feii\ emission bands in the Ark 120, finding that they do not
respond to the continuum variation. Thus the optical \feii-emitting
region may be heated by other mechanism than photoionization. These
results favor the shock heating for the optical \feii-emitting region,
but there are also difficulties. First, the amount of shock-processed
matter would probably be too large. Second, as \citet{kue} showed,
collisionally ionized models failed to match the shape of the optical
\feii-emission band. Third, the fact that there is no response to the
continuum variation for optical \feii\ emission bands can also be
interpreted as that the emitting region is too large to vary optical
\feii\ emission in observable timescales. Unless shocks are a viable
solution, the failure of the photoionization model simply indicates
that it is not predicting the correct heating rate, or that the
radiative transport calculations are not correct.

One possible cause disturbing classical photoionization models to
reproduce the observations may be the assumption that the \feii\
emission is isotropic. \citet{fer09} recently suggested that UV \feii\
lines are beamed toward a central source while optical \feii\ lines
are emitted isotropically. Then photoionization models can reproduce
the observed UV to optical \feii\ flux ratio if the \feii-emitting
clouds are distributed asymmetrically so that we mainly observe their
shielded faces. However, this needs special geometrical distributions
like Type II AGNs; a thick \feii-emitting gas surrounding the central
source with a substantial covering factor, and intervening between the
central region and our eyes. At the present time, it is not clear
whether or not photoionization models can reproduce the emission line
strengths other than \feii\ under such the situation. Much broader
exploration of photoionization model calculations is certainly
needed.

\subsection{Column density distribution inferred from \feiiopt/\feiiuv}
\begin{figure}
  \includegraphics[width=84mm]{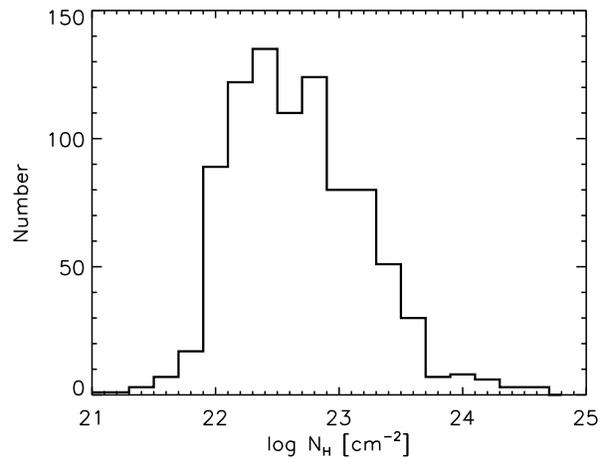}
  \caption{
    Estimated column density distribution. The average and the
    standard deviation of the distribution are $(\overline{\log N_H},
    \sigma_{\log N_H}) = (22.8, 0.5)$, providing observational support
    for the assumption adopted in \citet{mar09}.
  }
\label{fig:column_dist}
\end{figure}

Now we can roughly estimate column densities from the observed
\feiiopt/\feiiuv\ by using equation (\ref{eq:polyfit}). The estimated
column density distribution is shown in Figure
\ref{fig:column_dist}. An average and a standard deviation of the
distribution are found to be $(\overline{\log N_H}, \sigma_{\log N_H})
= (22.8, 0.5)$. \citet{mar09} has suggested that radiation pressure
does play an important role in BLR gas dynamics if column densities of
BLR clouds have intrinsic dispersion such as $(\overline{\log N_H},
\sigma_{\log N_H}) = (23.0, 0.5)$. Our results support that the
assumption adopted in \citet{mar09} is appropriate and that the
radiation pressure plays an important role in BLR clouds.

\subsection{Correlation between Eddington ratio and \feiiopt/\feiiuv}
\begin{figure}
  \includegraphics[width=84mm]{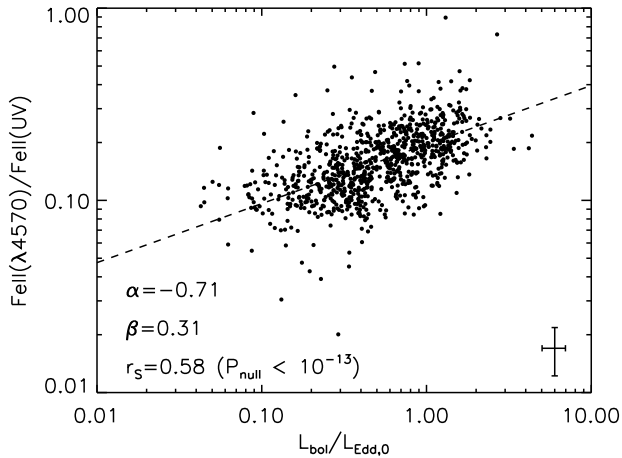}
  \caption{
    Eddington ratio vs. \feiiopt/\feiiuv\ flux ratio. Typical
    1$\sigma$ error is indicated at the lower right corner. The dashed
    line is the ``FITEXY.pro'' fit in the form of $\log y=\alpha +
    \beta \log x$. The values of $\alpha, \beta$ and Spearman's rank
    correlation coefficient $r_S$ are shown in the figure.
  }
\label{fig:fefe_edd}
\end{figure}

Figure \ref{fig:fefe_edd} shows the relation between \feiiopt/\feiiuv\ and the Eddington ratio. 
A positive correlation is seen. Linear regression analysis, using an
IDL procedure ``FITEXY.pro'' (cf. \citealt{pre}), gives the relation
as:
\begin{equation}
  \log \frac{\mbox{\feiiopt}}{\mbox{\feiiuv}} = -0.71 + 0.31 \log
  \frac{L_{bol}}{L_{Edd,0}} \label{eq:fefeedd}
\end{equation}
The Spearman's rank correlation coefficient\footnote{Spearman's rank
  correlation coefficient is non-parametric measure of correlation,
  that is, which assesses how well an arbitrary monotonic function
  could describe the relationship between two variables without making
  any other assumptions} for assessing the nonlinear correlation is
$r_S=0.58$. This means the probability of the null hypothesis that
there is no correlation is less than $10^{-13}$. Thus the correlation
between \feiiopt/\feiiuv\ and the Eddington ratio is real. This
implies that the column density increases with the Eddington ratio,
because \feiiopt/\feiiuv\ increases with the column density.

As was recently suggested by \citet{dong}, under the condition where
the BLR clouds are subject to the radiation pressure,
low-column-density clouds would be blown away by relatively large
radiation pressure at large $L_{bol}/L_{Edd,0}$, so that only
high-column-density clouds would be able to be gravitationally
bound. Figure \ref{fig:fefe_edd} is a supportive evidence for their
suggestion.

\subsection{Super-Eddington problem}
\begin{figure}
  \includegraphics[width=84mm]{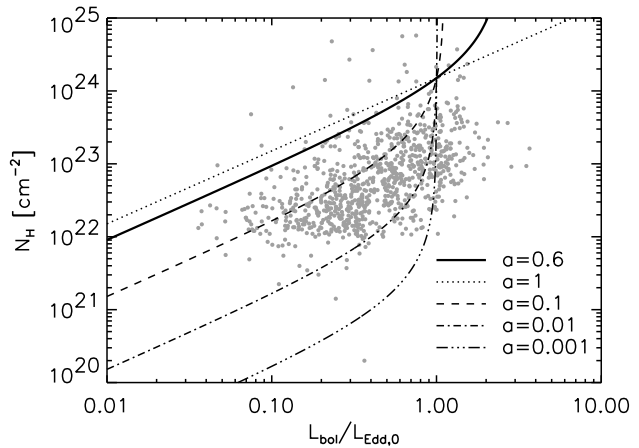}
  \caption{
    Estimated column densities vs. $L_{bol}/L_{Edd,0}$. Each line
    represents $L_{bol}=L_{Edd,rad}$. Thick solid: $a=0.6$ (adopted in
    \citealt{mar08}); dotted: $a=1$; dashed: $a=0.1$; dot-dashed:
    $a=0.01$; dot-dot-dot-dashed: $a=0.001$. Note that
    $a=L_{ion}/L_{bol}$. All these lines intersect at $N_H=1.5 \times
    10^{24}\ \mathrm{cm^{-2}}$ and $L_{bol}/L_{Edd,0}=1$, where
    $\sigma_T N_H=1$.
  }
\label{fig:edd_permit}
\end{figure}

Figure \ref{fig:edd_permit} plots our samples on $N_H -
L_{bol}/L_{Edd,0}$ plane. Each line represents $L_{bol}=L_{Edd,rad}$,
so that the lower region of the line corresponds super-Eddington
area. If we adopt ionizing photon fraction $a=0.6$ (i.e., thick solid
line in Figure \ref{fig:edd_permit}), which is an average value for
AGNs calculated by \citet{mar08}, almost all of our samples become
super-Eddington. This result can be interpreted in two ways: (i) the
conversion from \feiiopt/\feiiuv\ to column densities (i.e, equation
(\ref{eq:polyfit})) is wrong, or (ii) the adopted value of $a$ is
inappropriate.

In the case (i), since \feiiopt/\feiiuv\ is a function of the column
density and the temperature, the false in the conversion is attributed
to the assumed temperature. If we assume hot clouds such as $T_e >
10,000$ K, the corresponding column densities would become large,
resulting in the solution for this super-Eddington problem. However,
as already discussed, the previous studies favor cold clouds such as
$6,000 < T_e < 10,000$ K for \feii\ emission (cf. \citealt{col},
\citealt{joly}). Thus this interpretation seems to be inappropriate.

In the case (ii), as can be seen from Figure \ref{fig:edd_permit}, if
we adopt small values for $a$ such as 0.01, the majority of our
samples becomes gravitationally bound. This means that the fraction of
ionizing photons irradiating on \feii-emitting clouds is much less
than those on usual BLR clouds. Then it seems quite natural to
conclude that the \feii\ emission does not originate in the region
where usual emission lines such as H$\beta$ originate, but originate
in outer parts of BLR where the incident ionizing photon fraction
becomes as low as $a=0.01$. It is worth noting that from the studies
of \oi\ and \caii\ emission lines, \citet{mat08} also suggests that
\feii\ emission originates in outer parts of BLR.

\subsection{Eigenvector 1 in terms of the column density}

\citet{bg92} applied a principal component analysis to low-redshift
quasars and found that the principal component 1 (which is called
``Eigenvector 1'') links the strength of optical \feii\ emission and
the weakness of \oiii\ emission. After a while, \citet{bor} showed
that Eigenvector 1 is driven predominantly by an Eddington
ratio. However, the physical causes making up the Eigenvector 1 has
been left unknown.

Here we propose a physical interpretation of Eigenvector 1 in terms of
the column density. As discussed in the previous section,
small-column-density clouds would be driven away from the
line-emitting region by the radiation pressure at large Eddington
ratio, and only large-column-density clouds can be gravitationally
bound. Radiative transfer effects make the optical \feii\ emission
become large in such large-column-density clouds. On the other hand,
ionizing photons emitted from the central object are intervened by
these large-column-density clouds and thus have less probabilities of
ionizing photons reaching to NLR clouds, resulting in weak \oiii\
emission. In fact, Figure 10(t) in \citet{tzk} shows a negative
correlation between the \oiii/\hbeta\ and \feii(O1)/\feii(U1), which
is almost the same as our \feiiopt/\feiiuv, for 14 quasars. 

\section{Summary}

\begin{enumerate}

\item Analysis of the \feiiuv, \feiiopt\ and \mgii\ emission lines is
  performed for 884 SDSS quasars in a redshift range of $0.727 < z <
  0.804$.

\item We suggest that \feiiopt/\feiiuv\ can be an indicator of the
  column density of \feii-emitting clouds regardless of the excitation
  mechanism, i.e., photoionized or collisionally ionized clouds. From
  model calculations, we have confirmed this suggestion.

\item Our photoionization models underpredict \feiiopt/\feiiuv\ by a
  factor of 10, consistent with the preceding studies. Unless shocks
  are a viable heating mechanism, the failure of the photoionization
  model simply indicates that it is not predicting the correct heating
  rate, or that the radiative transport calculations are not
  correct. Ignoring anisotropy of UV \feii\ emission may be one of the
  causes.

\item The column density distribution estimated from \feiiopt/\feiiuv\
  is almost the same as the one suggested by \citet{mar09}, supporting
  that the radiation pressure does work on \feii-emitting clouds.

\item We also find a positive correlation between \feiiopt/\feiiuv\
  and the Eddington ratio, implying the links between the column
  density and the Eddington ratio.

\item We find that under the assumption of the ionization fraction
  $a=0.6$, almost all of our samples become super-Eddington. This
  problem can be cleared if the \feii\ emission originates in outer
  parts of BLR where the ionizing photon fraction becomes as low as
  $a=0.01$.

\item  We propose physical interpretation of 'Eigenvector 1' in terms
  of the column density. In the interpretation, the strength of the
  optical \feii\ emission results in the radiative transfer effects,
  while the weakness of the \oiii\ emission results in the reduction
  of ionizing photons in NLR caused by intervening large column
  density BLR clouds.

\end{enumerate}

\section*{Acknowledgments}
We thank the anonymous referee for providing us with very helpful
comments. This work was supported in part by Grant-in-Aid for JSPS
Fellows, Scientific research (20001003), Specially Promoted Research
on Innovative Areas (22111503), Research Activity Start-up (21840027),
and Young Scientists (22684005).

\end{document}